
_______________________________________________________________________________
\normalbaselineskip=20pt
\baselineskip=20pt
\magnification=1200
\hsize 16.0true cm
\vsize 22.0true cm

\def\lsim{\mathrel{\rlap{\lower4pt\hbox{\hskip1pt$\sim$}}
    \raise1pt\hbox{$<$}}}         
\def\gsim{\mathrel{\rlap{\lower4pt\hbox{\hskip1pt$\sim$}}
    \raise1pt\hbox{$>$}}}         

\def\ut#1{$\underline{\smash{\vphantom{y}\hbox{#1}}}$}
\def\overleftrightarrow#1{\vbox{\ialign{##\crcr
    $\leftrightarrow$\crcr
    \noalign{\kern 1pt\nointerlineskip}
    $\hfil\displaystyle{#1}\hfil$\crcr}}}
\long\def\caption#1#2{{\setbox1=\hbox{#1\quad}\hbox{\copy1%
\vtop{\advance\hsize by -\wd1 \noindent #2}}}}

\centerline{\bf COHERENT NUCLEAR DIFFRACTIVE}
\centerline{\bf PRODUCTION OF MINIJETS -}
\centerline{\bf ILLUMINATING COLOR TRANSPARENCY}
\vskip 12pt
\centerline{L. Frankfurt$^1$ and G.A. Miller}
\vskip 6pt
\centerline{\it Nuclear Theory Group, Department of Physics, FM-15}
\centerline{\it University of Washington, Seattle, Washington 98195}
\vskip 6pt
\centerline{and}
\vskip 6pt
\centerline{M. Strikman}
\vskip 6pt
\centerline{\it Department of Physics}
\centerline{\it Pennsylvania State University, University Park, PA 16802}
\vskip 6pt
\centerline{\bf Abstract}
\vskip 6pt
The cross section for coherent
nuclear processes in which an incident meson or photon
dissociates diffractively into two
jets of high relative transverse momentum  (between about 1 and 3 GeV/c)
is found to be proportional in QCD to $A^2$ in a wide kinematical region.
The leading correction to this is a significant positive term
$\propto A^{7/3}$ caused by nuclear
multiple-scattering. We find also that shadowing of hard
diffractive processes reappears as a leading twist effect
in the TeV energy range.

\vskip 16pt\noindent 1. Address after Feb. 1, 1993 Dep't of Physics, Tel Aviv
University, Ramat Aviv, Israel and Inst. of Nuclear Physics, St. Petersburg
, Russia.
\vfill\eject
\noindent I. \ut{Introduction}
The color fields emitted by closely separated systems of quarks and gluons are
cancelled for color singlet systems.  This effect of color neutrality CN
(or ``color screening") may lead to the suppression of initial
 or final state interactions [1](cf. the color transparency CT
phenomenon [2]) if a small sized system or point-like configuration PLC is
formed in a hard process.
For many two body hadronic processes, perturbative
QCD predicts that a PLC is formed at least
at extremely large momentum transfers [2-3].
Furthermore, an analysis of realistic models of hadrons hints that
a PLC may be formed at moderate momentum transfer $\sim 1 GeV/c$[4].
So far PLC are experimentally confirmed only for virtual
photons, and
a clear observation of CN and CT would be an important clue to some poorly
understood problems of dynamics of bound states in QCD [4,5].

The ideal situation to observe CN and CT effects would occur when it is
certain that PLC is formed, and when the energies are high enough so that the
expansion of the PLC [6,7] does not occur (the ``frozen" approximation is
valid).  The use of the 500 GeV/c
$\pi$ beam at the Fermi National Lab to
study coherent nuclear processes in which a pion diffractively dissociates
into a $q\bar q$ pair of high relative transverse momentum $\vec \kappa_t$
may present such an opportunity.  If the final
$q\bar q$ pair carries of all of
energy of the pion, the process involves only the
$q\bar q$ component of the light-cone wave function of the pion.
If it  is then possible to observe
coherent transitions in which the nucleus remains in its ground state, the
amplitude ${\cal M}(A)$
will be approximately proportional to
the nucleon number $A$.   This
$A^2$ variation in the cross section could supply a strong enough signal to
make the experimental observation of CT possible.
The jets we consider involve high (greater than about 1 GeV/c) but
not very high values of $\kappa_t$, less than
about 3 or 4 GeV/c (for currently accessible energies),
so we use the term minijets.

We focus on using a pion beam because experimental investigations
seem imminent [8,9]. However, a kaon or real photon beam could also be
employed, with expectations similar to those presented here.
Note also that pion diffraction to jets
could be  studied in the multi-TeV energy range using proton-proton
 colliders, if pion exchange is selected by observing the
highest energy $\Delta$'s or neutrons ($n$) produced at low $p_t$[10].
Indeed, for  the reaction:
$pp \rightarrow \Delta (n)  + p $ + 2 jets,
 the diffractively produced jets would appear in the
 central detector, so studying this reaction at FNAL may be feasible.

 The present searches for CT
using $(e,e'p)$[11] and $(p,pp)$ reactions[12] do not have
all of the advantages of
the hard $\pi$ diffractive processes. For example, the
 expansion of the PLC as it moves
through the nucleus is an important effect for those
experiments performed at much lower energies.

This paper
extends the original work of
Ref.[13]
and the naive constituent quark two-gluon exchange model of
Refs. [14-15] which focus on
the A-dependence of the
total cross section of diffractive processes.
We examine
coherent hard diffractive processes which select the PLC;
use the
dominance of gluon fusion diagrams, justified in QCD; and employ
QCD to constrain the pion wave function.

\noindent 2. \ut{Hard $\pi$-Nucleon Diffraction}
At high pion momentum $P_\pi$
(500 GeV), the $\pi$
breaks into a $q\bar q$ pair well before hitting the target, Fig.~1.
In that case,
the momentum transfer to the high momentum
$q\bar q$ system is transverse ($t\approx -q^2_t$), so the $q\bar
q$-nucleon interaction is essentially independent of the longitudinal
momentum of $q\bar q$ pair. The frozen approximation, in which
the transverse $q\bar q$ separation $b$ is a constant, should be valid
at such a high energy.
The observation of two final state jets carrying a total momentum very
close to $\vec P_\pi$ allows us to consider only $q\bar q$ components of
the pion, $\psi_\pi(x,\vec b)$.  The invariant amplitude ${\cal M}(N)$ for a
nucleon target is then given at large invariant energies $s$
by $${\cal
M}(N) = \int d^2b\psi_\pi(x,\vec b)\;{f(b^2)\over 2} e^{i\vec \kappa_t
\cdot\vec b},\eqno(1)$$ here
$f(b^2)$ is the forward $q\bar q$ scattering amplitude normalized
according to the optical theorem $Im\;f(b^2) = s\;\sigma(b^2)$, and
$t=0$. We neglect the small real part of $f(b^2)$.
The factor $e^{-i\vec \kappa_t\cdot\vec b}$ accounts for the final
hadronic wave function of
two  jets
with a high relative momentum.

The assumed
$b^2$ dependence of the interaction $\sigma(b^2)$ allows an evaluation of
Eq.~(1) in terms of the Fourier transform $\tilde\psi_{\pi}(x,\vec \kappa_t)$
by
expressing $b^2$ as $b^2 = -\nabla^2_{\kappa_t}$.  Then
$${\cal M}(N) = i\;{1\over 2}\;s\;\sigma(-\nabla^2_{\kappa_t})
\tilde\psi_\pi(x,
\vec \kappa_t)\;.\eqno(2)$$

\noindent 3. \ut{$q\bar q$ interaction- $\sigma(b^2)$}
At sufficiently small $b$ the two gluon exchange term is the
leading twist
effect, which is the dominant term at large s.
The QCD calculation of $\sigma_{PQCD}(b^2)$ [16]
involves a diagram similar to the gluon
fusion contribution to the nucleon sea-quark content
observed
in deep inelastic scattering.
One  calculates the box
diagram for large values of $\kappa_t$ using the wave function
of the pion
instead of the vertex for $\gamma^*\to q\bar q$.
  The result  is
$$\sigma_{PQCD}(b^2)\approx {4\pi^2\over 3}\;b^2\;
\alpha_s(\kappa^2_t)x_N\;G_N(x_N,2\kappa^2_t)
\equiv \sigma_0b^2/<b^2>,\eqno(3)$$
for small values of $b$. Here
$x_N = 2\kappa^2_t/s$, with $\kappa_t^2 \approx{1\over b^2}$.
and $G_N$ is the gluon distribution function in a
nucleon (or nucleus).

An evaluation of Eq.~(3) for a nucleon target
yields $\alpha_s x_N  G_N\approx 0.75$ implying $\sigma_0=$
 24 mb for $x_N \sim 10^{-2}$
and  $b^2\approx <b^2>$, which is
close to experimental value.
The evaluation of Eq.~(1)
requires a model for
$\sigma(b^2)$ for values of $b^2$ that aren't
very small. So the  higher twist corrections  neglected
above should be accounted for. A simple form that leads
to Eq.~(3) at small $b^2$ and also
roughly accounts for soft low momentum transfer physics is
$$\sigma(b^2) = {\sigma_0g\over\gamma}\left(1-e^{-\gamma b^2/<b^2>}\right),
\eqno(4)$$
where $<b^2>$ is the mean-square transverse separation $<b^2> = {2\over 3}
<r^2>\approx$ 0.24 fm$^2$.  For small $b^2$, Eq.~(4) reduces to Eq.~(3) but
with the additional factor $g$. We take it
between 1 and 3 to account for  uncertainties
in $\sigma_{pQCD}(b^2)$. These are
caused by the lack of precise knowledge
of $G_N$, the uncalculated
contribution of the  $q\bar q$-gluon component
(estimates using
dispersion sum rules indicate that this gives a
correction of about 30\%, V.Braun private communication).
The difference in
longitudinal  momenta  between the $q$ and $\bar q$ in the pion neglected
in deriving eq. (3) leads also to a 30\%
uncertainty.

\noindent 4. \ut{Non-perturbative model of $\tilde\psi_
\pi(x,\vec \kappa_t)$}
The approximate scale invariance of QCD in
one loop approximation shows that for large
$\kappa_t$,
$\tilde\psi_
\pi(x,\vec \kappa_t)\sim\;1/\kappa_t^2$
up to terms involving the  $log(\kappa_t)$,  which can be
calculated
using the renormalization group equations, see e.g. Refs.~3,17.

In practical estimates it is necessary to account for non-leading
powers of $1/\kappa_t^2$ (non-leading twist effects). We model them as
$$\tilde\psi^{(L^2)}_\pi(x,\kappa^2_t) = B(L^2){\phi(x)\over \kappa^2_t+\mu^2}
\;.\eqno(5)$$
 The constant $\mu^2$ is introduced to parameterize
poorly understood long distance effects.
The resulting wave function is defined at a non-perturbative
renormalization scale $L^2\approx 1\;GeV^2 $..
Note that the Fourier transform of $\psi_{\pi}^{(L)}(x,\kappa_t)$ leads to
$\psi_{\pi}(x,b)\propto K_0(\mu\;b)$,
where $K_0$ is the modified Bessel function of order zero.

The term $\phi(x)$ which is really a function of $\alpha_s(\kappa_t)$ also,
should tend to $\phi(x)=
\phi_{as}(x)
=\sqrt3 \;x(1-x)$ [18] at large $\kappa_t$.
The common wisdom is that at moderate $\kappa_t$ $\phi(x)$ is  far
from asymptotic value,more close
to  $\phi_{cz}(x) = 5\sqrt3\;x(1-x)(1-2x)^2$ [19].
The parameter
$\mu^2$ may  depend on $x$. However, we are mainly concerned with
the kinematic
region for which higher twist effects are not important, so
a detailed treatment of such effects is beyond the scope
of this paper. We
simply
treat $\mu^2$ as a constant and
use $\phi(x)=\phi_{as}(x)$
or $\phi_{cz}(x)$
in eq. (5).

The constant parameter $\mu$ is determined by choosing the
expectation value of $b^2$ in this wave function. We find
 $<b^2>={2\over 3\mu^2}$, so $\mu=0.33 GeV$ if $<b^2>=0.24 fm^2$,
corresponding to the measured pion size.
The $q\bar q$ component may have a
smaller size, and the sensitivity to $\mu^2$ is displayed in eq. (8).

The coefficient $B(L^2)$ can be obtained from the weak pion
decay constant[18]:
$$\int_0^1 dx \int^{L^2}d^2\kappa_t
{\tilde\psi^{(L^2)}_\pi(x,\kappa^2_t)\over 16\pi^3}=
{f_{\pi}\over 2 \sqrt{n_c}}
,
\eqno(6)$$
which determines
$B(L^2)= 16 \pi^2 f_{\pi}/ln (1+L^2/\mu^2)$ as a function of $L^2$, a
non-perturbative momentum transfer scale
which we take to be
$1 GeV^2$. The probability
that the pion is in a $q\bar q$ configuration is then
about 20\%, consistent
with upper bounds [18].  This bound is  evidently violated in the
constituent quark  model of pion used in [14-15].

Given $\tilde\psi^{(L^2)}_\pi(x,\kappa^2_t)$,  the parameters $\gamma$ and $g$
of $\sigma(b^2)$, eq.(4), may be determined. We
assume  that for soft processes the $q\bar q$ component of the
pion
 has the same cross section as the pion, 25 mb. This value can be reproduced
by a set of values of $\gamma$ and $g$. Values of
$\gamma$ from
0.1 to 1.0 can be used if $g$ varies between 1.2 and 2.8.

We wish to use the pion wave function at large values of $\kappa_t$, so
a QCD evolution of $\tilde\psi$ is needed.
But such effects provide small logarithmic corrections, which we
neglect in this first evaluation.

\noindent 5.  \ut{Leading Twist Dominance}
The elements needed to evaluate ${\cal M}(N)$ of
Eq. (1)  are now specified in
Eqs. (4) and (5).
The results (obtained with two sets of $\gamma,g$) displayed
in Fig.~2 show that ${\cal M}(N)\propto 1/\kappa_t^4$ for large values of
$\kappa_t$. These amplitudes have  a zero for $\kappa_t\approx 0.7 GeV/c$,
so our model is an oversimplification for such small
values of $\kappa_t$.

We are concerned with large values of $\kappa_t$, so
it is useful to evaluate  Eq.~2 using a Taylor series expansion for the
exponential of $\sigma(-\nabla^2_{\kappa_t})$.
Each Laplacian brings in an additional power of $\kappa_t^{-2}$, so
the leading term $\sim b^2 1/ \kappa_t^2=-4/ \kappa_t^4$. There
is a higher twist ${1/\kappa_t^6}$ term proportional
to $\gamma$ arising from $b^4$ term of
$\sigma(b^2)$ and another from the wave function proportional to $\mu^2$.
The evaluation of ${\cal M}(N)$ to order
of $1/\kappa_t^6$ yields
$${\cal M}(N)=-i {s\over 2} \;{\sigma_0g\over <b^2>}
{4\over \kappa^4_t}\left[
1+{2\mu^2\over \kappa^2_t}(6\gamma-2)\right]
\left\{B(L^2)\;\phi(x)\right\}\;.\eqno(7)$$
  The ${2\mu^2\over \kappa^2_t}$ term
 is small if $\kappa^2_t\gg {2\mu^2 (6\gamma-2)}
\approx
1 {\rm GeV}^2$ (for $\gamma=1$)
which is not a very large value. Thus ${\cal M}(N)$ is
well approximated by its leading twist value
for experimentally relevant kinematics; see
Fig. 2.

\noindent 6. \ut{Cross-section Estimate}
We compute the cross section from the invariant amplitude. The result
is
$${d^3\sigma_N\over dxdM^2_Jd^2P_{N_t}} = {1\over (2\pi)^4 16}{m_N|{\cal M}
_N|^2\over\sqrt{m_N^2+P^2_N}s^2}\eqno(8)$$
where $M^2_J = \kappa_t^2/x(1-x)$ is the mass squared of the jet
system, $x$ is the fraction of the pion momentum carried by the quark,
 $P_N$ is the magnitude of the three momentum,
 $M_N$ the mass, and
$P_{N_t}$ is the transverse momentum of the final nucleon.  To be specific
we make a numerical estimate using $t=0$.  Then for $\kappa_t^2\gg
1 GeV^2$
$$ {d^3\sigma_N\over dxdM^2_J\cdot d^2P_{N_t}} = 2.6\;{\rm GeV}^{-6}
\left({{\rm GeV}\over \kappa_t}\right)^{8}\phi^2(x)
,\eqno(9)$$
with $g=2 $ and $L=1 GeV$ .
For $\kappa_t$ = 2 GeV and at the maxima of $\phi(x)$  we find
${d^3\sigma_N\over dxdM^2_Td^2P_{N_t}} = {1.8(3.0)\times 10^{-3}\;{\rm
GeV}^{-6}}$ where the larger (smaller) value is obtained using $\phi_{cz}$
($\phi_{as}$). These values show
that the diffractive production of jets is a small but measurable cross
section.
There are substantial uncertainties in the
magnitude of ${\cal M}(N)$ caused by the uncertainty in taking into account
the higher twist effects in the pion wave function, in the value of
$\gamma$ in the developed model, see Fig. 2. However, the
normalization of the
wave function is fixed by Eq. (6) so that Eq.~(9)
seems to be
a reasonable estimate.

The t-dependence of this cross section can also be determined,
in the perturbative regime, from
the  two-gluon form factor of the nucleon.
An educated guess for this t-dependence, based on the data on
diffractive  photoproduction of $\rho$-mesons at large $Q^2$ [20]
and diffractive production of $F_s$ mesons in neutrino
scattering[21], is  $e^{bt}$ with $b$
in the range between 3.5 and 5 $GeV^{-2}$.

\noindent 7. \ut{Optical Approximation for Coherent Hard Diffraction}
We now compute the nuclear amplitude ${\cal M}(A)$. Eq. (1) for a nucleon
target applies at an energy large enough so that time dilation effects
prevent the expansion of the $q\bar q$ pair as it moves through the
target. For a nuclear target, expansion occurs unless[6,7]
$2P_{\pi}/( M_J^2-m_{\pi}^2)\gg R_A$,
where $M_J$ is the invariant mass of the $q\bar q$ system.
 $M_J^2= \kappa_t^2/x(1-x )\approx 4\kappa_t^2$, so
$2P_{\pi}/( M_J^2-m_{\pi}^2)\approx 12 Fm$ for $P_{\pi}=500 GeV/c$ and
$\kappa_t=2 GeV/c$.
Thus using light or medium sized nuclei is preferable to very
heavy targets.
We shall assume kinematics such that the expansion
can be neglected. Then ${\cal M}(A)$
is obtained by including
nuclear multiple scattering effects. The result
is
$${\cal M}(A) = i\;s\int d^2b\psi_\pi(x,b) e^{-i\vec\kappa_t\cdot\vec b}
\int d^2B
e^{i\;\vec q_t\cdot\vec B}
\left(1-e^{-{\sigma (b^2)\over 2} T_A(B)}\right).\eqno(10)$$
$B$ is the nuclear impact parameter, $t=-\vec q^{\;2}_t$
and,
$T_A(B)=\int^\infty_{-\infty}\rho_A(\sqrt{B^2+Z^2})dZ.$
A Wood-Saxon form is used for the nuclear density
$\rho_A(R)$.

The factor $f_A(b^2)$, with
$$f_A(b^2)\equiv \int d^2B\left(1-e^{-{\sigma(b^2)\over 2}T_A(B)}\right),\eqno
(11)$$
accounts for the $q\bar q$ multiple scattering series of interactions on
different nucleons.  For small $b^2$, $f(b^2) \approx {\sigma(b^2)\over
2}\int d^2B\;T(B) = {\sigma(b^2)\over 2}\;A$, and
${\cal M}(A)\propto A$.
The ``image" function $f_A(b^2)$ is very sensitive to the form of $\sigma
(b^2)$ for large values of $b$.  This is shown for $A$ = 40 in Fig.~3.
For Eq.~(4) the asymptotic value of $f_A$ at large $b$ is $\approx
\pi R^2_A$, while
for ordinary eikonal theory $f_A\approx \pi R^2_A$ for all $b$. Thus
eikonal theory
would lead to  ${\cal M}(A)\propto A^{2/3}$.

The evaluation of ${\cal M}(A)$ of Eq.~(10)
at $q_t=0$ yields the results of Fig. 4. The ratio
${\cal M}(A)/A{\cal M}(N)$ rises quickly from its value at 1 GeV/c and
overshoots unity by a substantial ammount. We can understand this
by again expanding in powers of
$\sigma(b^2 = -\nabla^2_{\kappa_t})$.
This gives a nuclear higher twist expansion:
$${\cal M}(A) \approx
 A{\cal M}(N)\left[1+4\;{\sigma_0g\over<b^2>\kappa^2_t}
\int d^2Be^{i\;\vec q_t\cdot\vec B}{T_A^2(B)\over A}\right]\;.\eqno(12)$$
The integral $\int d^2BT^2_A(B)\approx 0.26 A^{4/3}$, so the leading twist
correction grows rapidly, relative to the leading term, as $A^{1/3}$.
The numerical results display the accuracy of the approximation (12).

The use of eq.(10) corresponds to keeping some terms of infinite
order in $1/\kappa_t^2$, while neglecting other higher twist  terms of the
same $\kappa_t$ dependence.
This is justified because  the series we keep involves  terms of
order ${A^{1/3}\over \kappa_t^2}$, and $A^{1/3}$ is a large parameter.

We may obtain the nuclear cross section by replacing $N$ by $A$ everywhere in
Eq.~(8).  For small $t$ and for
sufficiently large $\kappa_t $, we find that
$${d^3\sigma_A\over dxdM^2_Jd^2P_{A_t}} = {d^3\sigma_N\over dxdM^2_Jd^2
P_{N_t}}\;F^2_A(t)A^2\;,\eqno(13)$$
where
$F_A(t)$ is the nuclear form factor.

This $A^2$ dependence is a strong
signal of CT.
The cross section for the coherent nuclear process is larger than for the
nucleon target by a factor of $A^2$ (or more),
but falls rapidly as t
decreases from 0. No other process has this behavior, so using this
property  it should be possible to observe the coherent process.

The higher twist terms provide a significant positive
contribution, so we investigate the
t-dependence of Eq. (12).
The relative importance of the
second-order (in $\sigma_0$) higher-twist term  increases as t increases. The
 second (higher twist) term in Eq.~(12) enters with a positive sign,
in contrast with the familiar second-order term of
elastic hA scattering. Thus one obtains
a broader diffractive peak than
for  elastic scattering, an effect especially important in
for the lightest
nuclear targets.
 This indicates another
feature of the onset of CT - narrowing of
the $q_t$ distribution with increase of the
transverse momentum of the quark jets.

Equations (13) and (9) allow simple
evaluations of cross sections.
Uncertainties in input parameters allow only
estimates  of the absolute values.
 But, the similarity of the two sets of high
$\kappa_t$ results of Fig. 4 implies that
the ratio
${\cal M}(A)/{\cal M}(N)$
is less model dependent than either amplitude.

It would  be useful to study $d^3\sigma_N$ and $d^3\sigma_A$ separately.
Those are proportional to $\phi^2(x)$ and the $x$ dependence may allow
a determination of $\phi(x)$. Indeed, the asymptotic wave function has a
maximum at x=1/2, while the CZ wave function vanishes.

\noindent 8. \ut{Gluon Shadowing}
At sufficiently large energies the $q\bar q$ interaction with nuclei is
corrected
by the effects of gluon shadowing.  To see this, simply observe that in
leading twist approximation Eq.~(3) is applies for a nuclear target (A).
So far we have used
$G_A = AG_N$. But   at  small enough $x_N=2\kappa_t^2/s$,
the interactions between emitted
gluons spread out in space
even when the quarks and
anti-quarks are closely separated. Then gluon shadowing occurs.
The results of Ref. 22 can be used to show that  gluon shadowing
is a relatively small correction for the present kinematics.
But
at TeV energies $x_N$ can be smaller than $10^{-3}$ and
the gluon shadowing should become much more important.
 The gluon shadowing effect may be even larger than for sea quarks observed
in deep inelastic scattering
because the
QCD coupling constant is larger for the octet representation
than for the triplet one.

\noindent 9. \ut{Non-coherent Nuclear Processes}
The experimental resolution may cause effects of nuclear excitation (to low
lying levels) to be included in the determination of $d^3\sigma_A$. We
examine the effects of two kinds of nuclear excited states.  If
quasielastic
nucleon knockout  dominates, one can use closure to sum all final nuclear
states.
This gives $d^3\sigma_A$ a term proportional to $A$, which
 should be distinguishable  from the
$A^{1/3}$ behaviour of incoherent diffractive processes
expected from the Glauber approximation.
The quasielastic knockout would occur at $q_t$ greater than the Fermi
momentum with
$q_0=q_t^2/2M_N$. Thus the search for a cross section proportional
to $A^2$ at low $q_t$ and  $A$ at higher $q_t$ provide two independent
ways to search for color transparency.

Inelastic scattering
to lower lying energy nuclear levels is generated mainly by an
operator proportional to ${\partial\over\partial R_A}\rho_A(R)$, e.g.
Ref.~23.
This surface peaked function leads to an $A^{2/3}$ behavior for ${\cal
M}(A)$,  distinguishable from the $A$ characteristic of
CT. However if the measurement
allows inelastic nuclear excitation to dominate one
could get a false signal that traditional approximations are
correct. Fortunately , the planned experiments [8,9]
employ kinematics such
that the energy transfer is practically
the same as the longitudinal momentum transfer
($q_0=q_3 -  (M_J^2 +q_t^2)/2P_{\pi}$).
This prevents significant excitation of collective levels which
requires $q_3\gg q_0$.

\noindent 10. \ut{Summary, Assessment and Future Possiblities}
At high energies the $A$ dependence of the nuclear processes  $\pi + A\to
q\bar q + A$  at small $(A^2)$ and at large $q_t$ ($A$)
may allow the unambiguous
identification of color
transparency. The determination of
the $\pi$ distribution function $\phi(x)$
may also be possible. These studies would help to resolve old
questions regarding the pion wave function, and the region where
pQCD is valid.

One difficulty is that the predicted
cross section falls as $1/\kappa_t^8$. So
it seems worthwhile to investigate also
nuclear diffractive photoproduction
of minijets, of
large $\kappa_t$ charmed hadrons. For large $\kappa_t$,
  using the known hadronic
wave function of the photon
 leads to a cross section that falls only as
$1/\kappa_t^6$.
Future TeV experiments
could also test the prediction that gluon shadowing effects
will be  important.

\noindent  \ut{Acknowledgements}
We thank the USDOE for financial support.
\vskip 12pt

\vfill\eject
\centerline{\bf References}
\vskip 12pt
\item{1.} F. E. Low, Phys. Rev. D12, 163 (1975);S. Nussinov Phys. Rev.
Lett 34,1286 (1975);

\item{2.} A.H.~Mueller in Proceedings of Seventeenth rencontre de
    Moriond, Moriond, 1982 ed. J Tran Thanh Van (Editions Frontieres,
Gif-sur-Yvette, France, 1982) Vol. I, p13;
S.J.~Brodsky in Proceedings of the Thirteenth
Int'l Symposium
on Multiparticle Dynamics, ed. W.~Kittel, W.~Metzger and A.~Stergiou (World
Scientific, Singapore 1982,) p963.

\item{3.} H.-N. Li and G. Sterman, Nucl. Phys. B381,129 (1992);
J. Botts and G. Sterman, Nucl. Phys.B325, 62 (1989)

\item{4.} L. Frankfurt, G.A. Miller, and M. Strikman, Comments on Nuclear
and Particle Physics 21,1(1992); Precocious Dominance of Point Like
Configurations,
in Press Nucl. Phys. A.

\item{5.} L. Frankurt and M. Strikman, Phys. Rep. 160,235 (1988)

\item{6.} G.R.~Farrar, H.~Liu, L.L.~Frankfurt \& M.I.~Strikman, Phys.
Rev. Lett. 61 (1988) 686.

\item{7.}
B.K.~Jennings and G.A.~Miller, Phys. Lett. B236 (1990) 209;
  B.K.~Jennings and G.A.~Miller, Phys. Rev. D 44
,692 (1991); Phys. Rev. Lett. 70,3619 (1992); Phys. Lett. B274 ,442 (1992).

\item{8.} D. Ashery, M. Moinester, private communication.

\item{9.} T. Ferbel, private communication to FS.

\item{10.} W.D.Walker, talk at the
 Boulder SSC workshop, July 1992.

\item{11.} A.S.~Carroll et al. Phys. Rev. Lett. 61 (1988) 1698;
A.S.~Carroll et al. BNL expt 850.

\item {12.}  SLAC exp. NE-18, R.D.McKeown and R. Milner, spokesmen.

\item{13.} G. Bertsch, S.J. Brodsky, A.S. Goldhaber, J. Gunion,
Phys. Rev. Lett.
\ut{47}, 297 (1981).

\item {14.}   Al.B.Zamolodchikov, B.Z.Kopeliovich, and L.I.Lapidus,
JETP Lett. 33,595 (1981).

\item {15.} B.Z. Kopeliovich
Professor Habilitation Dissertation,1987, unpublished; private communication.

\item{16.} B. Bl\"{a}ttel, G. Baym, L.L. Frankfurt, and M. Strikman,
Phys. Rev. Lett.71, 896 (1993).

\item{17.} S. Coleman, Chapt. 3 in ``Aspects of Symmetry", Cambridge
University Press, Cambridge 1985.
\item{18.} S.J. Brodsky, G.P. Lepage, Phys. Rev. D\ut{22}, 2157 (1980);
S.J. Brodsky, G.P. LePage, Phys. Scripta 23,945 (1981); S.J. Brodsky,
Springer Tracts in Modern Physics 100,81 (1982).

\item{19.} V.L. Chernyak and A.R. Zhitnitski, Phys. Rep. 112,173 (1984).

 \item{20.}  P.Amaudruz et al , Z. Phys.C54,239 (1992).
\item{21.}  A.E. Asratian et al, CERN-PPE/92-191 (1992).

\item{22.} L. Frankfurt, M. Strikman, S. Liuti, Phys. Rev. Lett. \ut{65},
1725 (1990).
\item{23.} A. DeShalit and H. Feshbach ``Theoretical Nuclear
Physics", John Wiley \& Sons New York 1974

\vfill\eject
\centerline{\bf Figure Captions}
\vskip 12pt
\item{Figure 1.} Illustrative diagram for hard diffractive production of
jets; other indicated only by  +....
The longitudinal momentum of the quark  is $xP_{\pi}$.
\vskip 12pt
\item {Figure 2.} Single  nucleon amplitude, ${\cal M}(N)$. Solid-exact
evaluation, Eq. (1) ; dashed- Higher twist expansion Eq.(8). The
normalization is arbitrary.
\vskip 12pt
\item{Figure 3.} $f_A(b^2)$ of Eq.~(11) for $A$ = 40.
\vskip 12pt
\item{Figure 4.} Ratios of nuclear to nucleon amplitudes. Solid- exact
evaluation of Eq. (10). Dashed - higher twist expansion Eq.(12).
\bye